\documentstyle[12pt,epsfig,rotating]{article}
\newcommand{\bce}{\begin{center}}
\newcommand{\ece}{\end{center}}
\newcommand{\beq}{\begin{equation}}
\newcommand{\eeq}{\end{equation}}
\newcommand{\bea}{\vspace{0.25cm}\begin{eqnarray}}
\newcommand{\eea}{\end{eqnarray}}

\newcommand{\ba}{\begin{array}}
\newcommand{\ea}{\end{array}}

\newcommand{\doublespace}{
    \renewcommand{\baselinestretch}{1.6}\large\normalsize}

\def\lsim{\mathrel{\rlap{\lower4pt\hbox{\hskip1pt$\sim$}}
    \raise1pt\hbox{$<$}}}	  
\def\gsim{\mathrel{\rlap{\lower4pt\hbox{\hskip1pt$\sim$}}
    \raise1pt\hbox{$>$}}}	  

\def\Pom{{\bf I\!P}}

\def\lsim{\mathrel{\rlap{\lower4pt\hbox{\hskip1pt$\sim$}}
    \raise1pt\hbox{$<$}}}         
\def\gsim{\mathrel{\rlap{\lower4pt\hbox{\hskip1pt$\sim$}}
    \raise1pt\hbox{$>$}}}         

\def\Pom{{\bf I\!P}}

  \advance\oddsidemargin  by -1in
  \advance\evensidemargin by -1in
  \advance\topmargin      by -0.5in
\def\lsim{\mathrel{\rlap{\lower4pt\hbox{\hskip1pt$\sim$}}
    \raise1pt\hbox{$<$}}}         
\def\gsim{\mathrel{\rlap{\lower4pt\hbox{\hskip1pt$\sim$}}
    \raise1pt\hbox{$>$}}}         

\def\Pom{{\bf I\!P}}
\def\beq{\begin{equation}}
\def\endeq{\end{equation}}
\def\arr{\begin{eqnarray}}
\def\endarr{\end{eqnarray}}
\makeindex

\doublespace

\begin{document}


\phantom{.}{\bf \Large \hspace{10.0cm} DFTT 77/95 \\
November 1995
\vspace{0.4cm}\\ }

\begin{center}
{\bf\sl \huge Excitation of open charm
and factorization breaking in rapidity gap events at HERA}
\vspace{0.4cm}\\
{\bf M.~Genovese$^{a,b}$,
N.N.~Nikolaev$^{c,d}$  and B.G.~Zakharov$^{d}$
\bigskip\\}
{\it
$^{a}$ Theory Division, CERN, CH-1211, Geneva 23, Switzerland
\medskip\\
$^b$ Dipartimento di Fisica Teorica, Universit\`a di Torino,\\
and INFN, Sezione di Torino, Via P.Giuria 1, I-10125 Torino, Italy
\medskip\\
$^{c}$IKP(Theorie), KFA J{\"u}lich, 5170 J{\"u}lich, Germany
\medskip\\
$^{d}$L. D. Landau Institute for Theoretical Physics, GSP-1,
117940, \\
ul. Kosygina 2, Moscow 117334, Russia
\vspace{1.0cm}\\ }
{\Large
Abstract}\\
\end{center}
We develop the pQCD description of diffraction excitation of
heavy flavours in DIS
and we derive the analytic formulas for the mass spectrum  
in leading log$m_f^{2}$.
The result illustrates
nicely non-factorization properties of the QCD pomeron. We predict
a very steep rise of the charm content of diffraction
dissociation of photons at small $x_{\Pom}$.
We evaluate the contribution of open charm to scaling violations
in the structure function of the pomeron.
 \bigskip\\

\begin{center}
E-mail: kph154@zam001.zam.kfa-juelich.de
\end{center}

\pagebreak

Following Ingelman and Schlein
\cite{Ingelman}, 
Regge factorization
\cite{Karen} is often applied  to
diffraction dissociation (DD) of (virtual) photons $\gamma^{*}
+p \rightarrow X +p'$ into states $X$ of mass $M$ (large rapidity
gap (LRG) events), which are reinterpreted
as a deep inelastic scattering (DIS) on pomerons
radiated by the target proton, endowing the pomeron with
the usual attributes of a particle such as the partonic structure
function $F_{2\Pom}(\beta,Q^{2})$ and the flux of pomerons
$\phi_{\Pom}(x_{\Pom})/x_{\Pom}$ in the proton (\cite{Ingelman,Regge}:
\arr
\left.(M^{2}+Q^{2})
{ d\sigma_{D} (\gamma^{*}\rightarrow X)
\over dt\,d M^{2} }\right|_{t=0} =
{  \sigma_{tot}(pp) \over 16\pi}
{4\pi^{2} \alpha_{em}
\over Q^{2}}
\phi_{\Pom}(x_{\Pom})
F_{2\Pom}(\beta,Q^{2})\, .
\label{eq:1}
\endarr
Here $Q^{2}$ is the virtuality of the photon, $W$ and $M$
are c.m.s. energy in the photon-proton and photon-pomeron
collision, $\beta =Q^{2}/(Q^{2}+M^{2})$ is the Bjorken
variable for the lepton-pomeron DIS
and $x_{\Pom}=(Q^{2}+M^{2})/(Q^{2}+W^{2})=x/\beta$ is
interpreted as the fraction of the momentum of the proton
carried away by the pomeron.

The Ingelman-Schlein model has never been derived from
a QCD analysis. Quite to the contrary, the unequivocal
conclusion from the QCD approach to DD is the non-factorization
of the QCD pomeron \cite{NZ92}: the $x_{\Pom}$
dependence of $d\sigma_{D}$ {\sl can not}
be reabsorbed entirely in the
$Q^{2}, \beta$ and flavour independent pomeron flux
function $\phi_{\Pom}(x_{\Pom})$, whereas the
$\beta, Q^{2}$ and flavour dependence of $d\sigma_{D}$
{\sl can not} be contained entirely
in the $f\bar{f}$ structure function of the pomeron.
Different aspects of the non-factorization in DD have
been discussed in \cite{NZ92,NZ94,NZsplit,GNZ95,NZ91}; the
non-factorizable colour dipole approach to DD \cite{NZ92,GNZ95}
is well known to provide very good quantitative description of the
HERA data on LRG events \cite{H1F2Pom,ZEUSF2Pom}.
Regarding the applicability of pQCD, the crucial
observation \cite{NZ92} is that in DD of transverse photons, the
$q\bar{q}$ pairs have a, $Q^{2}$-independent, typical transverse
size $\sim 1/m_{f}$. This allows to quantify the factorization
breaking in DD into heavy flavours on a more quantitative basis
than for the (predominantly nonperturbative) DD into light
flavours. The demonstration of this factorization breaking
is the subject of the present paper.

One of the main points of the present communication
is a derivation of
the $\beta$ dependence of the pQCD factorization
scale,
\beq
q_{0}^{2}\sim m_{f}^{2}(1+{Q^{2}\over M^{2}})={m_{f}^{2}\over 1-\beta}
\,.
\label{eq:2}
\endeq
This $\beta,m_{f}$ dependence of the pQCD factorization scale
drives the flavour and $\beta$ dependence
of the flux of pQCD pomerons in the proton and
nicely illustrates the breaking of Regge factorization
for the QCD pomeron: each and every flavour brings in 
a new, and explicitly $\beta$-dependent,
flux function into the menagerie of pomeron fluxes
in the proton. We predict a very steep rise of the
charm 
abundance in DD towards small
$x_{\Pom}$, which is a distinctive feature of the QCD approach
to DD as compared to the Regge factorization models.
We present also an evaluation of the contribution
of open charm excitation to the counterintuitive rise with $Q^{2}$
of the pomeron structure function at large values of $\beta$.
Finally,  we comment on the even more dramatic factorization breaking in
DD of longitudinal photons, where
the pQCD factorization scale $q_{0}^{2}\sim {1\over 4}Q^{2}$
entails the $Q^{2}$-dependent "flux of pomerons" which can
be related to the $Q^{2}$ dependence of the gluon structure
function of the proton (for the similar situation in DD into
dijets see \cite{NZsplit}).

DD into open heavy flavour,
$X=q\bar{q}$, is described by diagrams of Fig.~1.
It dominates at $M^{2}\sim Q^{2}$ considered here.
The relevant formalism is described in detail in
\cite{NZ92,NZsplit}. The mass of
the state $X$ is given by
$
M^{2}=(m_{f}^{2}+k^{2})/z(1-z)\, ,
$
where $m_{f}$ is the quark mass, $\vec{k}$ is
the transverse momentum of
the quark with respect to the $\gamma^{*}$-pomeron collision
axis  and $z$ is 
the fraction of light--cone momentum
of the photon carried by the (anti)quark.
Other useful kinematical
variables are
$
\varepsilon^{2}=z(1-z)Q^{2}+m_{f}^{2}$ and
\beq
q^2=
k^{2}+\varepsilon^{2}=(k^{2}+m_{f}^{2}){M^{2}+Q^{2}\over M^{2}} \,
\label{eq:3}
\endeq

After the standard leading log$\kappa^{2}$ resummation,
the cross sections of the forward ($t=0$)
DD of (T) transverse and (L)
longitudinal photons
takes the compact form \cite{NZ92,NZsplit}
\arr
\left.{d\sigma_{T} \over
dM^{2}dk^{2}dt}\right|_{t=0}={\pi^{2}\over 6}
e_{f}^{2}
 \alpha_{em}\alpha_{S}^{2}(q^{2}) \cdot {m_{f}^{2}+k^{2} \over
 M^{3}\cos\theta \sqrt{M^{2}-4m_{f}^{2}} }
\left\{ \left(1-2{k^{2}+m_{f}^{2}\over M^{2}}\right)
\Phi_{1}^{2} + m_{f}^{2}\Phi_{2}^{2}
\right\}  \, ,
\label{eq:4}
\endarr
\arr
\left.{d\sigma_{L} \over
dM^{2}dk^{2}dt}\right|_{t=0}={\pi^{2}\over 6}
e_{f}^{2} \alpha_{em} Q^{2}
\alpha_{S}^{2}(q^{2})
\cdot {(m_{f}^{2}+k^{2})^{3} \over
M^{7}\cos\theta \sqrt{M^{2}-4m_{f}^{2}}}
\Phi_{2}^{2}  \, .
\label{eq:5}
\endarr
Here $\theta$ is the quark production angle
with respect to the $\gamma^{*}$-pomeron collision axis,
\arr
\Phi_{1}=\int {d\kappa^{2}\over \kappa^{4}}
f(x_{\Pom},\kappa^{2})
\left[{k\over k^{2}+\varepsilon^{2}}-
{k\over \sqrt{a^{2}-b^{2}}}+
{2k\kappa^{2} \over a^{2}-b^{2}+a\sqrt{a^{2}-b^{2}}}\right]\, ,
\label{eq:6}
\endarr
\beq
\Phi_{2}=\int {d\kappa^{2}\over \kappa^{4}}
f(x_{\Pom},\kappa^{2})
\left[{1\over \sqrt{a^{2}-b^{2}}} -
{1\over k^{2}+\varepsilon^{2}}
\right] \, ,
\label{eq:7}
\endeq
$a=\varepsilon^{2}+k^{2}+\kappa^{2}$,
$b=2k\kappa$
and $f(x_{\Pom},\kappa^{2})=\partial
G(x_{\Pom},\kappa^{2})/\partial \log \kappa^{2}$
is the unintegrated gluon structure function of the target proton.
In the derivation of the pQCD factorization scale we follow
the analysis \cite{NZsplit}: At small $\kappa^{2}\lsim q^{2}$
the expression in the square brackets in the integrand of
(\ref{eq:6}) equals $2k\epsilon^{2}\kappa^{2}/q^{6}$ and tends to
a constant value at $\kappa^{2}\gsim q^{2}$. Then,
in (\ref{eq:6}) and (\ref{eq:7}) one has a logarithmic
$\kappa^{2}$ integration with $q^{2}$ being the upper limit
of integration. Consequently, $q^{2}$ emerges as the pQCD factorization
scale (it has already
been used as such in the running strong coupling
$\alpha_{S}(q^{2})$
in (\ref{eq:4}) and (\ref{eq:5})) and
to the leading log$q^{2}$,
\arr
\Phi_{1}=
{2kM^{4}[Q^{2}(k^{2}+m_{f}^{2})+M^{2}m_{f}^{2}]
\over
(Q^{2}+M^{2})^{3}(k^{2}+m_{f}^{2})^{3} }
G(x_{\Pom},q^2)
\label{eq:8}
\endarr
\beq
\Phi_{2}=
{M^{4}[(k^{2}+m_{f}^{2})(M^{2}-Q^{2})-2m_{f}^{2}M^{2}]
\over
(Q^{2}+M^{2})^{3}(k^{2}+m_{f}^{2})^{3} }
G(x_{\Pom},q^2)\, .
\label{eq:9}
\endeq
Notice a zero of the $d\sigma_{L}$ at
$
(k^{2}+m_{f}^{2})(M^{2}-Q^{2})=2m_{f}^{2}M^{2}$.
Equations (\ref{eq:3}),
(\ref{eq:8}) and
(\ref{eq:9}) exhaust the derivation of the mass spectrum.

Subtleties of DD in QCD are clearly seen
already for real photons ($Q^{2}=0$), where
\arr
\left.{d\sigma_{T} \over
dM^{2}dk^{2}dt}\right|_{t=0}=
{1\over 6}\pi^{2}e_{f}^{2}
 \alpha_{em} \alpha_{S}^{2}(m_f^{2}+k^{2})
G^{2}(x_{\Pom},m_{f}^2+k^{2})
m_{f}^{2}\nonumber\\
\times \left[( k^{2}+m_{f}^{2})(1-{8m_{f}^{2} \over M^{2}})+
{8m_{f}^{4}\over M^{2}}
\right]
{1\over  M^{3}(k^{2}+m_f^{2})^{4}\cos\theta \sqrt{M^{2}-4m_{f}^{2}}}
  \,.
\label{eq:10}
\endarr
Evidently, the $k^{2}$-integrated
 $d\sigma_{T}/dm^{2}$ is dominated by the contribution
from
$k^{2} \lsim m_{f}^{2}$. Consequently, to the leading log$m_{f}^{2}$
one can take the $k^{2}$-independent factorization scale
$q_{0}^{2}= m_{f}^{2}$. (The dominance of
of the contribution from $k^{2} \lsim m_{f}^{2}$
in $\sigma_{T}$ holds also at large $Q^{2}$, entailing
the pQCD factorization scale $q_{0}^{2}$ of Eq.~(\ref{eq:2})).
We find
\arr
\left.{d\sigma_{T} \over
dM^{2}dt}\right|_{t=0}=\pi^{2}e_{f}^{2}
 \alpha_{em} \alpha_{S}^{2}(m_{f}^{2})
G^{2}(x_{\Pom},m_{f}^2)\times
\left\{\begin{array}{ll}
{ \sqrt{M^{2}-4m_{f}^{2}}
\over  96 m_{f}^{7}},& ~~{\rm if}~~M^{2}- 4m_{f}^{2}\lsim m_{f}^{2}\\
{1\over  12m_{f}^{2}M^{4}},&~~{\rm if}~~M^{2}\gg 4m_{f}^{2}\\
\end{array}
\right.
\,.
\label{eq:11}
\endarr
The $M^{2}$ and $x_{\Pom}$
dependence in (\ref{eq:11}) do factor and
one may try to
reinterpret the
$x_{\Pom}$ dependent factor
as the pomeron flux function
\beq
\phi^{(f\bar{f})}_{\Pom}(x_{\Pom})=
\left({G(x_{\Pom},m_{f}^2)\over
G(x_{0},m_{f}^2)}\right)^{2}\, ,
\label{eq:12}
\endeq
subject to the normalization $\phi^{(f\bar{f})}_{\Pom}(x_{0}=0.03)=1$
\cite{GNZ95}. The explicit flavour dependence in (\ref{eq:12})
breaks the Regge factorization, hence
the universal flux of QCD pomerons does not exist.

Consider now the DIS regime of large $Q^{2}$.
For transverse photons,
the expression in the curly braces in Eq.~(\ref{eq:4}) is a
function of only $\beta$ and $Q^{2}$ and can be a basis for the
definition of the $f\bar{f}$ pomeron structure function
$F_{\Pom}^{(f\bar{f})}(\beta,Q^{2})$. The factor
$\alpha_{S}^{2}(q_{0}^{2})=\alpha_{S}^{2}(m_{f}^{2}(1-\beta)^{-1})$
also is a function of $\beta$ only
and as such can be reabsorbed into the
$F_{\Pom}^{(f\bar{f})}(\beta,Q^{2})$.
The $x_{\Pom}$ dependence comes entirely from 
$G^{2}(x_{\Pom},q_{0}^2)=G^{2}(x_{\Pom},m_{f}^{2}(1-\beta)^{-1})$ and
here by virtue of the QCD scaling violations
the $x_{\Pom}$ and $\beta$ dependences are inextricably
entangled, which
breaks the Regge factorization (\ref{eq:1})  explicitly. 
One can think of a generalized factorization
at best, leaving the pomeron
flux factor to depend explicitly on $\beta$ and on flavour:
\beq
\phi_{\Pom}^{(f\bar{f})}(x_{\Pom},\beta)=
\left({G(x_{\Pom},m_{f}^{2}(1-\beta)^{-1})\over
G(x_{0},m_{f}^{2}(1-\beta)^{-1})}\right)^{2}\, .
\label{eq:13}
\endeq
Because $q_{0}^{2}$ rises with
$\beta$, the
pQCD is applicable better at smaller $M$; for the 
specific case of
exclusive vector meson production see \cite{NNZ94}.

At this point,
a brief digression on the still more striking, and different,
Regge factorization breaking in the longitudinal cross section is
in order \cite{GNZlong}.
Substituting (\ref{eq:8}) into (\ref{eq:5}), one readily
finds
\arr
\left.{d\sigma_{L} \over
dM^{2}dk^{2}dt}\right|_{t=0}={1\over 6}
\pi^{2}e_{f}^{2} \alpha_{em} Q^{2}
\alpha_{S}^{2}(k^{2}+\varepsilon^{2})
G^{2}(x,k^{2}+\varepsilon^{2})\nonumber\\
\times {M[(k^{2}+m_{f}^2)(M^{2}-Q^{2})-2m_{f}^{2}M^{2}]^{2}
\over
\cos\theta (Q^{2}+M^{2})^{6}(k^{2}+m_{f}^{2})^{3}
\sqrt{M^{2}-4m_{f}^{2}} }\,
\label{eq:14}
\endarr

which decreases with $k^{2}$ only as $k^{-2}$.
While $\sigma_{T}$ is dominated
by the contribution from $k^{2}\lsim m_{f}^{2}$,
for $\sigma_{L}$ the dominant contribution
comes from large $k^{2} \sim {1\over 4}M^{2}-m_f^{2}$.
This has
two major implications: First,  $\sigma_{L}$ has the higher twist
$Q^{2}$ dependence $\sigma_{L} \propto \sigma_{T}/Q^{2}
\propto 1/Q^{4}$ \cite{NZ91}. Second,
in $\sigma_{L}$ the factorization scale
$
q_{0}^{2}\sim {1\over 4} Q^2
$, the $x_{\Pom}$ and $Q^{2}$ dependence in $d\sigma_{L}$ are
inextricably entangled and the generalized flux of
pomerons does explicitly depend on $Q^{2}$,
$\phi_{\Pom}^{L}(x_{\Pom}) \propto
G^{2}(x_{\Pom},{1\over 4}Q^{2})$, rather than on $\beta$ and flavour
in the case of $d\sigma_{T}$.

The exceptional case
is the triple pomeron region of $\beta \ll 1$
dominated by DD into the
$q\bar{q}g..$ states, where the conditions of the Regge factorization
are fulfilled at $Q^{2}\gsim
3$\,GeV$^{2}$ for all the flavours simultaneously and the
corresponding flux function $f_{\Pom}(x_{\Pom})$ is flavour
blind. This conclusion readily follows from an analysis of the
colour dipole content of the triple-pomeron coupling in
ref.~\cite{GNZA3Pom}.

Hereafter we focus on DD of transverse photons
into $c\bar{c}$ states, which dominates the open charm excitation
in LRG events
at $\beta\gsim $0.1-0.2. For open charm the factorization scale
(\ref{eq:2}) is still not large, we evaluate the open charm cross
section in the colour dipole gBFKL formalism described in
detail in \cite{NZ92,NZ94,GNZ95}.
In Fig.~2 we show our results for the total cross section of
diffraction excitation of $c\bar{c}$ pairs.
The evaluation of this cross section requires the $t$-integration;
one can argue that for the heavy flavour excitation the
diffraction slope $B$ is smaller than
the light flavours one ($B_{el}\approx 10$\,GeV$^{-2}$).
In the following we use $B_{c\bar{c}}=6$\,GeV$^{-2}$ (for instance,
see \cite{NZZslope}).
The closer analysis of Eqs.~(\ref{eq:4}) suggests a simple
interpolation $\sigma_{T} \propto 1/(Q^{2}+4m_{f}^{2})$
between the real photoproduction ($Q^{2}=0$) and
DIS ($Q^{2}\gg 4m_{f}^{2}$) at fixed value of the natural
variable $x_{Pom}=(Q^{2}+4m_{c}^{2})/(W^{2}+Q^{2})$. This
scaling is demonstrated in Fig.~2 where we
plot $(Q^{2}+4m_{c}^{2})\sigma_{T}^{(c\bar{c})}$ as
a function of the Bjorken variable $x$ and
$x_{Pom}$.

Notice the very steep rise, $\sigma_{T}\propto x_{Pom}^{-\epsilon}$,
at $x_{Pom}\lsim 10^{-3}$, with the exponent
$\epsilon \approx 0.72$ which is
very close to the asymptotic gBFKL prediction
$\epsilon = 2\Delta_{\Pom}=0.8$ (for the prediction of the
precocious onset of the BFKL behaviour in the
charm structure function of the proton see \cite{NZDelta}).
In Fig.~3 we show the flux function
$\phi^{(c\bar{c})}_{\Pom}(x_{\Pom})$
defined for the above
integrated $c\bar{c}$ excitation cross section
at large $Q^{2}$, which
is dominated by the contribution from $\beta \sim 0.5$.
For the  region $0.1 \lsim x_{\Pom}
\lsim 10^{-4}$ accessible at HERA, the convenient parameterization is
\beq
\phi^{(c\bar{c})}_{\Pom}(x_{\Pom})=
\left({x_{o}\over x_{\Pom}}\right)^{p_{1}}
\left({x_{\Pom}+p_{3}\over x_{o}+p_{3}}\right)^{p_{2}}
\label{eq:15}
\endeq
with $ p_{1}=0.7233, p_{2}=0.3939, p_{3}=2.377\cdot10^{-3}$,
which is different from the flux function $\phi_{\Pom}(x_{\Pom})$
for the valence light-flavour component of the pomeron
($ p_{1}=0.569, p_{2}=0.4895, p_{3}=0.153\cdot10^{-3}$) and
the flux function $f_{\Pom}(x_{\Pom})$ for the sea (triple-pomeron)
component of the  pomeron ( $ p_{1}=0.741, p_{2}=0.586,
p_{3}=0.8\cdot10^{-3}$)
\cite{GNZ95}.
The charm content of
DD is predicted to rise by one order
of magnitude from $x_{\Pom}=0.01$ to $x_{\Pom}=0.0001$
(see also Fig.~6).

Now we focus on the $Q^{2}$ and $\beta$ dependence of the open
charm excitation. The variation of the factorization scale
(\ref{eq:2}) at $\beta\lsim 0.5$ is marginal, but
at $1-\beta \ll 1$ the
factorization-breaking
$\beta$ dependence of the generalized flux function
(\ref{eq:15}) is quite strong. The distortion factor
\beq
\Gamma(\beta)=
{\phi_{\Pom}^{(c\bar{c})}(x_{\Pom},\beta) \over
\phi_{\Pom}^{(c\bar{c})}(x_{\Pom},\beta=0.5)}
\label{eq:16}
\endeq
presented in
Fig.~4, shows how  the shape of the $\beta$
distribution varies with  $x_{\Pom}$
in defiance of the Regge factorization.
Here we  have evaluated $\Gamma(\beta)$ 
using
the GRV gluon structure function \cite{GRV}, for other
parameterizations of parton densities
the results for the factorization
breaking will be very similar.

Despite having discredited the
very concept of the pomeron structure
function, we can not help but use this language to make the contact
with what has unfortunately become a common presentation of the
experimental data on DD. The $t$-integrated cross section
measured at HERA  can be represented as
\arr
(M^{2}+Q^{2})
{ d\sigma_{D} (\gamma^{*}\rightarrow X)
\over d M^{2} } =
{  \sigma_{tot}(pp) \over 16\pi B_{el}}
{4\pi^{2} \alpha_{em}
\over Q^{2}}F_{D}(x_{\Pom},\beta,Q^{2})\,
\label{eq:17}
\endarr
with the valence light ($q\bar{q}$) and valence charm ($c\bar{c}$)
decomposition of the non-factorizing "diffractive structure
function" 
(we omit the negligible $b\bar{b}$ contribution,
neglect the marginal difference between fluxes for the
$s\bar{s}$ and $u\bar{u},d\bar{d}$ excitation and limit ourselves
to the transverse structure function.)
\beq
F_{D}(x_{\Pom},\beta,Q^{2})=
\phi_{\Pom}(x_{\Pom})
F_{\Pom}^{(q\bar{q})}(\beta,Q^{2})+
\phi_{\Pom}^{(c\bar{c})}(x_{\Pom},\beta)
F_{\Pom}^{(c\bar{c})}(\beta,Q^{2})+
f_{\Pom}(x_{\Pom})
F_{\Pom}^{(sea)}(\beta,Q^{2})\, .
\label{eq:18}
\endeq
The result \cite{GNZ95} for light flavours,
$
F_{\Pom}^{(q\bar{q})}(\beta)\approx 0.27\beta(1-\beta),$
 evaluated with the diffraction
slope $B_{el}=10$\,GeV$^{-2}$,
provides an excellent description of the experimental data
\cite{H1F2Pom,ZEUSF2Pom}.
In Fig.~5 we show our predictions for the $\beta$ dependence of
the $c\bar{c}$ component of $F_{D}(x_{\Pom},\beta,Q^{2})$
at $x_{\Pom}=0.001$.
Because $M^{2}\gsim 4m_{c}^{2}$, the
$F_{D}^{(c\bar{c})}(x_{\Pom},\beta,Q^{2})$ vanishes
at
$\beta>\beta_{c}={Q^{2}\over Q^{2}+4m_{c}^{2}}\,.$
The impact of this threshold effect on
$ F_{D}^{(c\bar{c})}(\beta,Q^{2})$ was for the first time
discussed in \cite{NZ92}.
Fig.~5
updates Fig.~10 of ref.~\cite{NZ92};
in the present calculation we use the more modern
gBFKL dipole cross section of \cite{NZHera,NNZ94}.

In Fig.~6 we show the threshold effect
due to opening of the charm production in
$F_{D}(x_{\Pom},\beta,Q^{2})$ considered as a function
of $Q^{2}$ at different values of $\beta$.
Here we neglect the possible scaling violations
in the light flavour contribution $F_{D}^{(q\bar{q})}
(x_{\Pom},\beta,Q^{2})$ (see below). In the domain of
$0.35 \lsim \beta \lsim 0.8$ and
$5\lsim Q^{2}\lsim 200$\,GeV$^{2}$
of the current HERA experiments, the predicted
threshold rise
of $F_{D}(x_{\Pom},\beta,Q^{2})$ is quite strong
and must be observable. The error bars of the
presently available HERA
data \cite{H1F2Pom,ZEUSF2Pom}
on $F_{D}(x_{\Pom},\beta,Q^{2})$ are still too
large for the observation of the charm threshold effect.
Fig.~6 clearly
demonstrates the rise of the charm content of DD towards
small $x_{\Pom}$: at the typical $\beta \sim 0.5$ and
 $x_{\Pom}=10^{-2}$ not shown here
the charm contribution and the threshold effect are
$\sim 3\%$, which rises to $\sim 9\%$ at
$x_{\Pom}=10^{-3}$ and $\sim 25\%$ at $x_{\Pom}=10^{-4}$,
cf. different fluxes in Fig.~3.
At  even smaller $x_{\Pom}$, not accessible at HERA, the
charm content of DD levels off.
The strong $x_{\Pom}$-dependence of the charm content
of DD is a non-negotiable consequence of pQCD; such a
$x_{\Pom}$-dependence is
absent in Regge models \cite{Capella,Golec,Gehrmann}.
Neither  works \cite{Capella,Golec} discuss the strong
impact of the charm threshold on the $Q^{2}$ dependence
of DD at large $\beta$. One of the two models
of DD into open
heavy flavour discussed in \cite{Gehrmann} assumes a
pointlike pomeron-quark coupling; in this model too
the charm content does not depend on $x_{\Pom}$. Such
a pointlike pomeron-quark coupling is not born out
in our pQCD approach.

Several more comments on our results are in order:

First, for heavy flavours the results for the
$\beta$ distribution are exact in contrast to the nonperturbative
DD into light flavours, where the $\beta$ dependence
is not pQCD calculable. Fig.~5 shows that at very large $Q^{2}\gg
4m_{c}^{2}$, in a broad range of $\beta$,
the $\beta$ dependence of
$F_{D}^{(c\bar{c})}(x_{\Pom},\beta,Q^{2})$
follows the approximation $\propto \beta(1-\beta)$ 
fairly well.
In terms of the mass spectrum, it corresponds to
$d\sigma_{D}^{(c\bar{c})}/dM^{2} \sim M^{2}/(Q^{2}+M^{2})^{3}$. The
departures from this law were noticed already in \cite{NZ92}
(see Fig.~3 in \cite{NZ92}). Indeed, in a very narrow region
of $\beta \rightarrow 1$, i.e., $4m_{c}^{2} \ll M^{2} \ll Q^{2}$, from
Eqs.~(\ref{eq:4}),(\ref{eq:8}),(\ref{eq:9}) one readily
finds (modulo to the pQCD scaling violations)
$d\sigma_{D}^{c\bar{c}}/dM^{2} \sim M^{4}/(Q^{2}+M^{2})^{4}$
and  $F_{D}^{c\bar{c}}(x_{\Pom},\beta,Q^{2}) \propto (1-\beta)^{2}$.
Although this observation is hardly of practical significance,
the law $F_{D}(x_{\Pom},\beta,Q^{2}) \propto (1-\beta)^{2}$
will be applicable also to the light flavour excitation
in a very narrow domain for $\beta \rightarrow 1$, in
which the factorization scale (\ref{eq:2}) is sufficiently
large for the pQCD applicability
(for a hint at such a $\beta$ dependence
see also Fig.~3 in \cite{NZ92}).

Second, the factorization breaking
and the $\beta$-dependent pQCD factorization scale
have certain implications for
the $Q^{2}$-evolution of $F_{D}^{c\bar{c}}(x_{\Pom},\beta,Q^{2})$,
which remains
one of the open issues in the theory of DD.
In \cite{NZ94} it was shown that
an approximate GLDAP evolution is recovered
in the double leading-log$Q^{2}$, leading-log${1\over \beta}$
approximation. Here we wish to comment that Eq.~(2) implies
that DD at large $\beta$ is dominated by excitation of
$q\bar{q}$ pairs with the transverse size $r_{q\bar{q}}$
which decreases with $\beta$,
\beq
r_{q\bar{q}}^{2} \propto {1\over m_{f}^{2}}(1-\beta)\, .
\label{eq:19}
\endeq
At large values of the
Bjorken variable $\beta$ the parton distributions
decrease with $Q^{2}$ for the radiation of gluons. Evidently,
the radiation of gluons by a colour singlet system of size
$r_{q\bar{q}}$ requires the condition $Q^{2}r_{q\bar{q}}^{2} \gg 1$.
Then, Eqs.~(\ref{eq:2}),(\ref{eq:19}) hint at the possibility
that the larger $\beta$ the larger $Q_{0}^{2} \propto (1-\beta)^{-1}$
is needed for the onset of the conventional pattern of the
$Q^{2}$ evolution of $F_{D}^{c\bar{c}}(x_{\Pom},\beta,Q^{2})$
at $Q^{2}> Q_{0}^{2}$.

Third, the predicted
steep rise of the charm excitation cross section
has a non-negligible impact on the $x_{\Pom}$ dependence
of $F_{D}(x_{\Pom},\beta,Q^{2})$. Consider the
$x_{\Pom}^{-\epsilon}$ approximation of this quantity in
the vicinity of $x_{\Pom}\sim 10^{-3}$ typical for the current
HERA experiments.
For excitation of light flavours $q=u,d,s$,
our prediction \cite{GNZ95}
for the flux function $\phi_{\Pom}(x_{\Pom})$
corresponds to $\epsilon(uds)
\approx 0.15$, which
is close to the
soft pomeron model predictions \cite{Regge,Capella,Golec,Gehrmann}.
For the charm excitation we predict $\epsilon(c) \approx 0.72$
at $x_{\Pom} \lsim 10^{-3}$. Although the abundance of
charm is numerically small,
the steeply rising contribution of the charm to
$F_{D}(x_{\Pom},\beta,Q^{2})$
substantially renormalizes upwards the exponent $\epsilon$
for the total DD cross section: at
$\beta \sim 0.5$ and $Q^{2}$ above the charm
threshold our crude estimate is
$\epsilon(uds+c)\approx 0.20$.
For comparison, in
the triple pomeron region of $\beta \ll 1$ our prediction \cite{GNZ95}
for the flux function $f_{\Pom}(x_{\Pom})$ corresponds to
$\epsilon \approx 0.32$. Evidently, the estimates for these
exponents depend on $\beta$ and on the range of $x_{\Pom}$ considered,
see Figs.~3 and 4.

To summarize, we have presented the pQCD derivation of the mass
spectrum and cross section of DD into open charm in DIS
at HERA. Our results unequivocally demonstrate strong
breaking of the Regge factorization in diffraction
dissociation of photons,
which has already
been advocated by two of the authors in a previous work
\cite{NZ92}. We predict a very steep
rise of the charm content of DD towards small $x_{\Pom}$,
which is a distinctive consequence of pQCD and
can be tested at HERA. We predict strong $x_{\Pom}$
dependence of the $\beta$ distribution in open charm
production at $\beta\rightarrow 1$, which also can be
tested in the HERA experiments.
We  have found a substantial charm threshold effects in the
$Q^{2}$ dependence of the diffractive structure function
at large $\beta$
which must be observable with higher statistics data
at HERA. This charm threshold effect
leads to a counterintuitive rise of the
large-$\beta$ diffractive structure
function with $Q^{2}$, which is stronger at smaller $x_{\Pom}$.
Our finding of  the $\beta$-dependent pQCD factorization
scale casts  shadow on the GLDAP description
of the $Q^{2}$-evolution of the diffractive structure
function at large $\beta$.
\medskip\\
{\bf Acknowledgments:} B.G.Zakharov thanks J.Speth for the
hospitality at the Institut f\"ur Kernphysik, KFA, J\"ulich.
This work was partly supported by the INTAS grant 93-239 and
the Grant N9S000 from the International Science Foundation.
\pagebreak

\end{document}